## Investigation of UV Absorbers on Venus Using the 283 and 365 nm Phase Curves Obtained From Akatsuki

Y. J. Lee[1] , A. García Muñoz[2] , A. Yamazaki[3,4] , M. Yamada[5], S. Watanabe[6] , and T. Encrenaz[7]

[1]Zentrum für Astronomie und Astrophysik, Technische Universität Berlin, Berlin, Germany, [2]AIM, CEA, CNRS, Université Paris-Saclay, Université de Paris, Gif-sur-Yvette, France, [3]Institute of Space and Astronautical Science (ISAS/JAXA), Sagamihara, Japan, [4]Graduate School of Science, University of Tokyo, Tokyo, Japan, [5]Planetary Exploration Research Center (PERC), Narashino, Japan, [6]Space Information Center, Hokkaido Information University, Ebetsu, Japan, [7]LESIA, Observatoire de Paris, PSL University, CNRS, Sorbonne Université, Université de Paris, Meudon, France

**Abstract** The so-called unknown absorber in the clouds of Venus is an important absorber of solar energy, but its vertical distribution remains poorly quantified. We analyze the 283 and 365 nm phase curves of the disk-integrated albedo measured by Akatsuki. Based on our models, we find that the unknown absorber can exist either well mixed over the entire upper cloud or within a thin layer. The necessary condition to explain the 365 nm phase curve is that the unknown absorber must absorb efficiently within the cloud scale height immediately below the cloud top. Using this constraint, we attempt to extract the $SO_2$ abundance from the 283 nm phase curve. However, we cannot disentangle the absorption by $SO_2$ and by the unknown absorber. Considering previous $SO_2$ abundance measurements at midinfrared wavelengths, the required absorption coefficient of the unknown absorber at 283 nm must be more than twice that at 365 nm.

**Plain Language Summary** There is an unknown absorber in the clouds of Venus. It absorbs solar energy effectively at ultraviolet (UV) and blue wavelengths, but its vertical location, either above or below the cloud top level (about 70 km altitude), remains unclear. This uncertainty affects our understanding of the vertical deposition of solar energy in the atmosphere. We investigate the vertical distribution of the unknown absorber using the dependence of the full-disk brightness on the scattering direction (the Sun-Venus-spacecraft angle) at 365 nm, with data from JAXA's Akatsuki spacecraft over 3 years. We find that the unknown absorber could exist in the entire cloud, or as a thin layer near but below the cloud top. Using these constraints on the vertical distribution of the unknown absorber, we analyze the 283 nm full-disk brightness. At this shorter wavelength, the $SO_2$ gas and the unknown absorber are both effective absorbers. We attempt to quantify the $SO_2$ abundance, and find that the brightness dependence on the scattering direction alone is insufficient to separate the contribution from the two absorbers at 283 nm. Further analysis with spectral phase curve observations will better define the $SO_2$ abundance.

## 1. Introduction

The 80% of solar energy incident on Venus is reflected back to space by the thick and highly reflective clouds that cover the entire planet (Moroz et al., 1985; Tomasko, Smith, et al., 1980). This reflected radiation has a strong directional dependency, which has been used to retrieve the microphysical properties of the cloud aerosols (Arking & Potter, 1968; García Muñoz et al., 2014; Lee et al., 2017; Mallama et al., 2006; Markiewicz et al., 2014; Petrova et al., 2015; Shalygina et al., 2015). The reflectivity of Venus is high at visible wavelengths and starts to decrease from the blue toward the ultraviolet (UV) (Barker et al., 1975). The absorption is caused by the $SO_2$ gas at $\lambda < 320$ nm (Esposito, 1980; Marcq et al., 2020; Zasova et al., 1981) and an unidentified secondary absorber (Pollack et al., 1980; Zasova et al., 1981). The latter is usually referred to as the "unknown absorber" (Titov et al., 2018) (hereafter, "UA"). The UA is reported to have its maximum absorption at 340 nm (Pérez-Hoyos et al., 2018) and becomes less absorbing toward shorter wavelengths (Marcq et al., 2020; Pérez-Hoyos et al., 2018), where it overlaps with $SO_2$ absorption.

The UA is responsible for ∼50% of the solar heating rate at the cloud top level (Crisp, 1986; Lee, Titov, et al., 2015). A recent 365 nm space-based data analysis revealed that the UA abundance has varied by a factor of 2 over the last decade (Lee, Jessup, et al., 2019), showing that such a change in the UA abundance and associated solar heating rate could alter the strength of the Hadley circulation. This may affect the super-rotation at the cloud top level atmosphere, explaining the observed long-term variations of wind speeds (Lee, Jessup, et al., 2019).

The vertical distribution of the UA, so far poorly constrained, also affects the solar energy deposition in the atmosphere. Tomasko, Doose, et al. (1980) reported that the solar energy deposition occurred mainly in the upper cloud (>60 km altitude) but not in the middle-to-low cloud (48–60 km). Pollack et al. (1980) and Esposito (1980) argued that the UA is located below the cloud top level to explain the observed phase angle dependence of disk-resolved UV contrasts. Kawabata et al. (1980) concluded from their polarimetric data that the small-size aerosols above the clouds (upper haze) cannot contain the UA. However, Molaverdikhani et al. (2012) argued for a well-mixed layer of the UA above the cloud top level. Lee, Immamura, et al. (2015) suggested that the vertical distribution of the UA varied over ∼5 years of Venus Express observations. They showed that initially the UA was present above the cloud top but that later was found below. In previous spectral data analyses, the UA has been assumed to be well mixed with the small-size aerosol particles (Marcq et al., 2020; Pérez-Hoyos et al., 2018).

Here, we explore whether the uncertain vertical distribution of the UA may be resolved by analyzing Akatsuki's UV images of the full planetary disk collected over 3 years. We use simultaneous 283 and 365 nm brightness measurements to attempt to extract the absorption of the UA at 283 and 365 nm, and the $SO_2$ abundance. We use ground-based measurements of global mean $SO_2$ abundance (Encrenaz et al., 2019) to compare and constrain further our findings.

## 2. Observations

The UV camera (UVI) on board Akatsuki has monitored the Venus dayside since December 2015, acquiring images at 283 and 365 nm (Yamazaki et al., 2018). The Akatsuki spacecraft has a highly elliptical equatorial orbit with an orbital period of ∼10 days (Nakamura et al., 2016). Except for a few hours during pericenter passage, UVI obtains global disk images at a phase angle less than 155°. In this study, we analyzed such full-disk images acquired from December 2015 to January 2019. We rejected images with an insufficient disk coverage within the field-of-view of UVI, for example, when the apparent size of Venus becomes too large, or when the disk location is too off-centered. We also rejected images that contained known artifacts. The total number of selected images are 5,805 at 283 nm and 5,840 at 365 nm; the complete list of them is given by Lee et al. (2020). We calculated the disk-integrated albedo $A_{\text{disk-int}}$, as the following (Sromovsky et al., 2001),

$$A_{\text{disk-int}}(\alpha, \lambda, t) = \frac{\pi}{\Omega_{\text{Venus}}(t)} \frac{d_{\text{V-S}}(t)^2 F_{\text{Venus}}(\alpha, \lambda, t)}{S_\odot(\lambda)}, \tag{1}$$

where $\lambda$ is the filter's effective wavelength, $t$ is the observation time, $\alpha$ is the Sun-Venus-spacecraft phase angle (°), $d_{\text{V-S}}(t)$ is the distance between the Venus and the Sun (AU) at the time of observation $t$, $F_{\text{Venus}}(\alpha, \lambda, t)$ is the disk-integrated flux of Venus (Wm$^{-2}$ μm$^{-1}$), $\Omega_{\text{Venus}}(t)$ is the solid angle of Venus as viewed from Akatsuki, and $S_\odot(\lambda)$ is the solar irradiance at 1 AU (W m$^{-2}$ μm$^{-1}$), calculated for the transmittance functions of each filter. The calculation of $A_{\text{disk-int}}$ is described in the supplementary information (Text S1). $A_{\text{disk-int}}(\alpha = 0°, \lambda)$ is the "geometric albedo" at wavelength $\lambda$.

All data points of $A_{\text{disk-int}}$ as a function of $\alpha$ are shown in Figure 1. We also show the mean phase curve over the 3 years. Short-term variations exist on time scales ranging from a few days to months (Lee et al., 2020), indicated as standard deviations in the same plot. The size of the phase angle bins for the mean phase curve goes from 1° at $\alpha < 30°$ to 5° at $30° \leq \alpha < 140°$, and 2° at $\alpha \geq 140°$. This helps capture the glory at small phases (García Muñoz et al., 2014; Lee et al., 2017; Markiewicz et al., 2014), and the decreasing $A_{\text{disk-int}}$ at large $\alpha$, while saving model calculation time (Section 3).

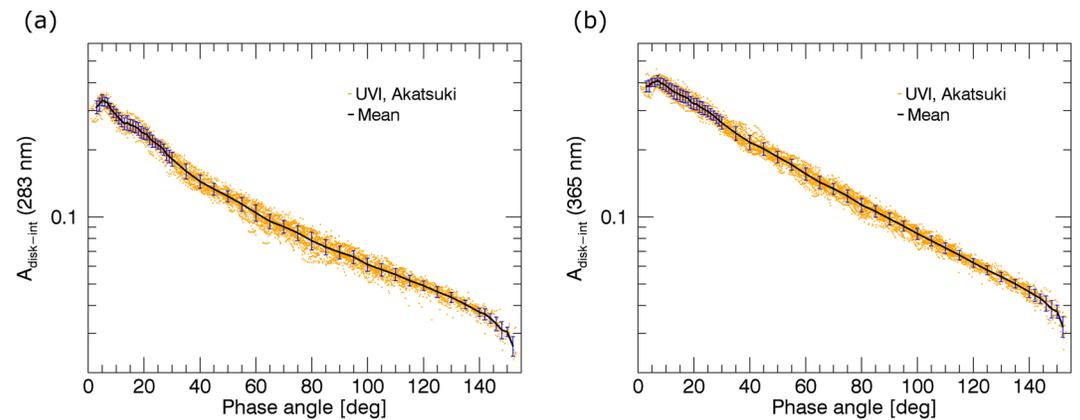

**Figure 1.** Observed disk-integrated albedo at (a) 283 nm and (b) 365 nm as a function of phase angle. Orange dots are the individual measurements obtained by Akatsuki. Black curves are the mean values. Blue error bars are standard deviations. See the text for details.

## 3. Model Descriptions and Atmospheric Conditions

### 3.1. Model Description

We fit the observed mean phase curves with model $A_{disk-int}$ curves obtained by solving the multiple scattering radiative transfer problem with a Preconditioned Backward Monte Carlo (PBMC) algorithm (García Muñoz, 2015; García Muñoz & Mills, 2015). We used a 1 km vertical grid to describe the atmosphere of Venus. We assumed a black surface below the cloud layer (48 km altitude), after confirming that this assumption does not change our results and saves calculation time (Figure S1). To produce each synthetic phase curve, we used $10^5$ photons, which was deemed a fine balance between calculation efficiency and accuracy.

### 3.2. Assumed Atmospheric Conditions

We assumed a uniform cloud over the entire disk. The cloud vertical structure above 60 km altitude is reasonably consistent with that used in previous studies, that is, 70 km altitude for the cloud top level (a unity optical thickness, $\tau = 1$, at 365 nm) and 4 km cloud scale height (Figure S2) (Haus et al., 2014; Ignatiev et al., 2009; Lee et al., 2012; Sato et al., 2020). Two modes of cloud aerosols were assumed: mode 1 with $r_{eff} = 0.43$ $\mu m$ and $\nu_{eff} = 0.52$ (Pollack et al., 1980) and mode 2 with $r_{eff} = 1.26$ $\mu m$ and $\nu_{eff} = 0.76$ (Lee et al., 2017). The size of mode 2 was inferred from the glory observation of UVI (Lee et al., 2017). We calculated the particles' scattering phase functions using the Lorenz-Mie code by Mishchenko et al. (2002) and a gamma distribution of particle sizes. The refractive indices of 75% $H_2SO_4$-$H_2O$ cloud aerosols were taken from Hummel et al. (1988). The cross sections of each mode vary by less than 2% from 365 to 283 nm. This means that the extinction introduced by the mode 1 and mode 2 is effectively the same at both wavelengths investigated here.

The ratio of extinction coefficients of mode 1 and mode 2 was assumed to be 1:1 at 365 nm. In such conditions, the particle number density ratio of mode 1 over mode 2 is ~320 (Figure S2b), according to the ratio of extinction cross sections for mode 1 over mode 2 at 365 nm. This number density ratio is close to the reported ~400 (Luginin et al., 2016). As smaller ratios were used in other past works (Haus et al., 2014; Pérez-Hoyos et al., 2018), we also explored the impact of a different extinction coefficient ratio, 0.5:1. The resulting differences in brightness are visible at small or large phase angles but not on the overall phase curve shape (Figure S3). Our main conclusions are not very sensitive to the extinction coefficient ratio (Figures 2 and S4) because the UA in our model is considered embedded in both cloud aerosol modes over the specified range of altitudes (Section 3.3). The corresponding extinction coefficient of the aerosols at 283 nm is calculated using the number density and the cross section at 283 nm for each particle mode.

Atmospheric pressure and temperature profiles at low latitudes were taken from Seiff (1983). Rayleigh scattering was calculated for the 96.5% $CO_2$ and 3.5% $N_2$ mixture (Sneep & Ubachs, 2005). The absorption cross section of $SO_2$ was taken from Vandaele et al. (2009) at 283 nm and ignored at 365 nm. We assumed that the $SO_2$ volume mixing ratio decreases exponentially with altitude with a 3 km scale height (Encrenaz et al., 2019; Marcq et al., 2020), and adjusted its abundance (ppbv) at the cloud top level. This enables the comparison between our results and those from Encrenaz et al. (2019) in the same manner as done in Marcq et al. (2020). Our prescribed 3 km scale height for the $SO_2$ vertical distribution is somewhat different from the increasing volume mixing ratio above the cloud top level inferred observationally by Vandaele et al. (2017). We therefore compared the inversion of vertical abundance above the cloud top and a constant $SO_2$ abundance over altitudes (Figure S5b), and found that the three $SO_2$ abundance profiles produced phase curves with nearly undistinguishable shapes (Figure S5a). Cloud top ozone, localized at high latitudes (>50°) (Marcq et al., 2019), is not considered in this study because of the small portion of the high latitudinal area in the fully illuminated equatorial view, 13%, and consequently its small global area averaged abundance, ~1 ppbv.

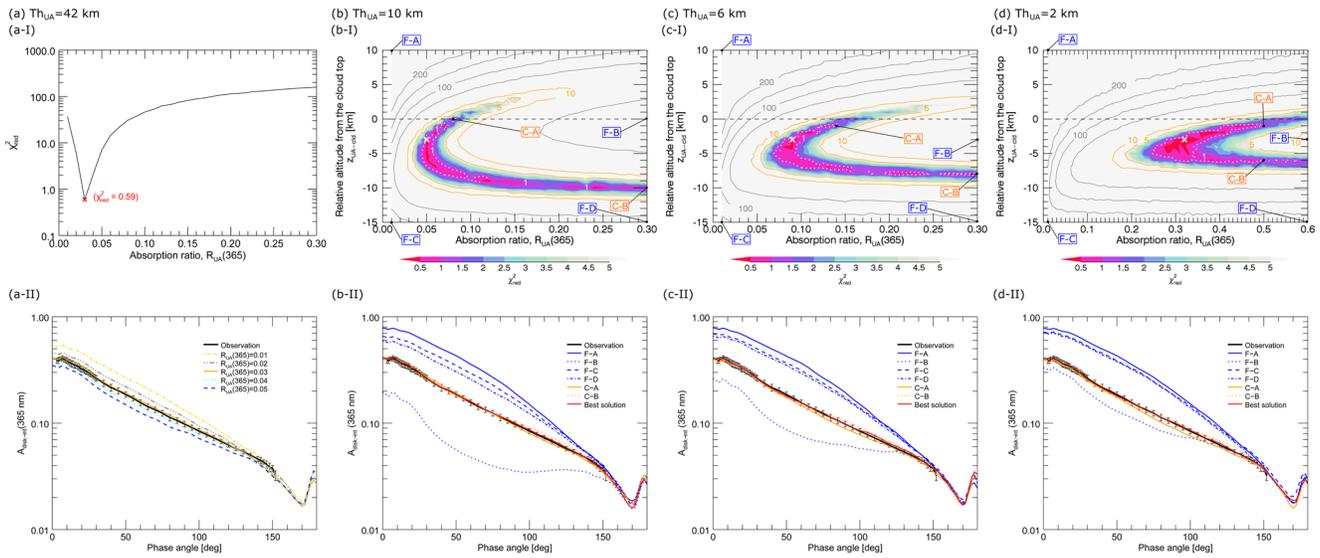

**Figure 2.** Comparison of the observed and calculated phase curves at 365 nm. $Th_{UA}$ varies from 42 (entire cloud) to 2 km, from left to right columns. The upper panels show $\chi^2_{red}$ as a function of $R_{UA}(365)$ (a), and both $R_{UA}(365)$ and $z_{UA-cld}$ (b–d). Note the range of $R_{UA}(365)$ in (d), which is upto 0.6. The dashed lines indicate the cloud top level. The white dotted lines indicate $\chi^2_{red}=1$. The area with the values beyond the range of the color bar ($\chi^2_{red}>5$) is filled with light gray. Orange contours have a different interval than gray contours. The lower panels compare phase curves for selected cases, marked in the upper panels of (b–d): the best solution is for the minimum $\chi^2_{red}$ ("×", $\chi^2_{red}=0.12$–0.19), false cases (from F-A to F-D), and comparably good solutions (C-A and C-B). Their exact conditions and $\chi^2_{red}$ values are listed in Table S3.

### 3.3. Free Model Parameters: $SO_2$ and the UA

In our synthetic phase curve calculations, the vertical distribution of the UA and the abundance of $SO_2$ are described through five free parameters, listed below. We incorporate the UA into the mode 1 and mode 2 particles by reducing the particles' single scattering albedo (SSA), even though in reality both particles absorb negligibly at the two specified wavelengths. This method of reducing the SSA of mode 1 particles to consider the effect of the UA has been used before (e.g., Crisp, 1986). Here, we apply this method for both modes to explore the impact of the vertical distribution of the UA. Thus, in our model we refer to aerosols to mean the combination of mode 1 and mode 2 particles, which include the UA at the prescribed altitude range. The five free parameters are:

- The thickness of the UA layer: $Th_{UA}$
- The altitude of the UA layer: $z_{UA-cld}$, which is the middle point between the top boundary altitude of the UA layer ($z_{UA, top}$) and the bottom boundary altitude of the UA layer ($z_{UA, bot}$)
- The ratio of UA absorption relative to aerosol extinction at 365 nm: $R_{UA}(365)$
- The ratio of UA absorption relative to aerosol extinction at 283 nm: $R_{UA}(283)$

- The $SO_2$ abundance in (ppbv) at the cloud top

We experiment with a number of configurations, as follows. (1) $Th_{UA}$ varies from 42 km (entire cloud) to 2 km (thin layer). (2) $z_{UA-cld}$ (also for $z_{UA,top}$ and $z_{UA,bot}$) varies with respect to the cloud top level, as above + or below − in steps of 1 km. (3) $R_{UA}(365)$ and (4) $R_{UA}(283)$ define the absorption of UA in terms of a ratio to the cloud extinction coefficient ($km^{-1}$). The scattering coefficients for the aerosols become $(1 − R_{UA}) \times$ Extinction coefficient. Then, the SSA for the cloud aerosols becomes $(1 − R_{UA})$, that is, SSA = Scattering/Extinction = $(1 − R_{UA})$. $R_{UA}$ is nonzero only over the range of altitudes, where the UA is prescribed with (1) and (2) and we vary $R_{UA}$ in steps of 0.01 to explore its effect. The scattering coefficients of the aerosols and Rayleigh scattering of the $CO_2$-$N_2$ atmosphere are used to calculate the scattering matrix of the atmosphere following the usual summation rules. (5) The $SO_2$ abundance is varied in steps of 0.02 in a logarithmic scale, that is, $d(\log_{10}(SO_2)) = 0.02$, from 0.1 to 1,000 ppbv.

We first investigate the phase curve at 365 nm because it is affected by only one active absorber (UA), while there are two active absorbers at 283 nm (UA and $SO_2$). We determine the combinations of $Th_{UA}$, $z_{UA-cld}$, and $R_{UA}(365)$ that are consistent with the observations at 365 nm. After that, with the best fitting configurations of $Th_{UA}$ and $z_{UA-cld}$, we attempt to determine $R_{UA}(283)$ and the $SO_2$ abundance at 283 nm.

## 4. Results

We determine a goodness of fit with a reduced chi-square ($\chi^2_{red}$):

$$\chi^2_{red} = \frac{1}{(N-1)} \sum_{i=1}^{N} \frac{(O_i - C_i)^2}{\sigma_i^2}, \qquad (2)$$

where $N$ is the number of phase angle bins ($N = 56$), $O_i$ is the mean observed $A_{disk-int}$, $\sigma_i$ is the associated standard deviation, $C_i$ is the calculated value, and $i$ denotes each phase angle bin from 3° to 152° (Table S2). We emphasize that $\sigma_i$ is not representative of the signal-to-noise ratio, which is much better than that for each data point. As seen below, the best fits result in $\chi^2_{red}$ notably less than one.

### 4.1. 365 nm Results

We first explore the configuration in which the UA spans the entire cloud ($Th_{UA} = 42$ km, $z_{UA,top} = +20$ km, $z_{UA,bot} = −22$ km). In this case, the 365 nm phase curve simulation has a single free parameter, $R_{UA}(365)$, and $R_{UA}(365) = 0.03$ fits the observation best ($\chi^2_{red}=0.59$) (Figure 2a). Based on this configuration, we experiment with reducing the top boundary of the UA ($z_{UA,top}$) from +20 to −10 km. Similarly, we experiment with increasing the bottom boundary of the UA ($z_{UA,bot}$) from −22 to +10 km. The resulting $\chi^2_{red}$ contours (Figures S6a and S6b, respectively) show that a good fit is achieved only when $z_{UA,top} \geq −5$ km or $z_{UA,bot} < −5$ km. We conclude that a vertical distribution of UA extending over the entire upper cloud provides a satisfactory fit to the observations as long as there is enough UA absorption within 5 km below the cloud top.

We next explore the cases of $Th_{UA}$ ranging from 10 to 2 km (Figures 2b–2d) and thus UA distributions over layers of varying finite size. For each $Th_{UA}$, the free parameters are $R_{UA}(365)$ and $z_{UA-cld}$. The configuration of minimum $\chi^2_{red}$ is marked with the "×" symbol in the upper panels (Figure 2I). Selected phase curves are compared in the lower panels (Figure 2II).

For $Th_{UA} = 10$ km (Figure 2b) the best solution corresponds to $R_{UA}(365) = 0.05$ and $z_{UA-cld} = −3$ km ($\chi^2_{red} = 0.17$). The $\chi^2_{red} \leq 1$ area indicates that $z_{UA-cld}$ should be between 0 and −10 km (C-A and C-B) and that the UA must absorb within the first few kilometers below the cloud top. At $z_{UA-cld} = −10$ km (UA at 55–65 km), the sensitivity to $R_{UA}(365)$ becomes weak even for $R_{UA}(365) > 0.10$ because the UA is located too deep to have an impact.

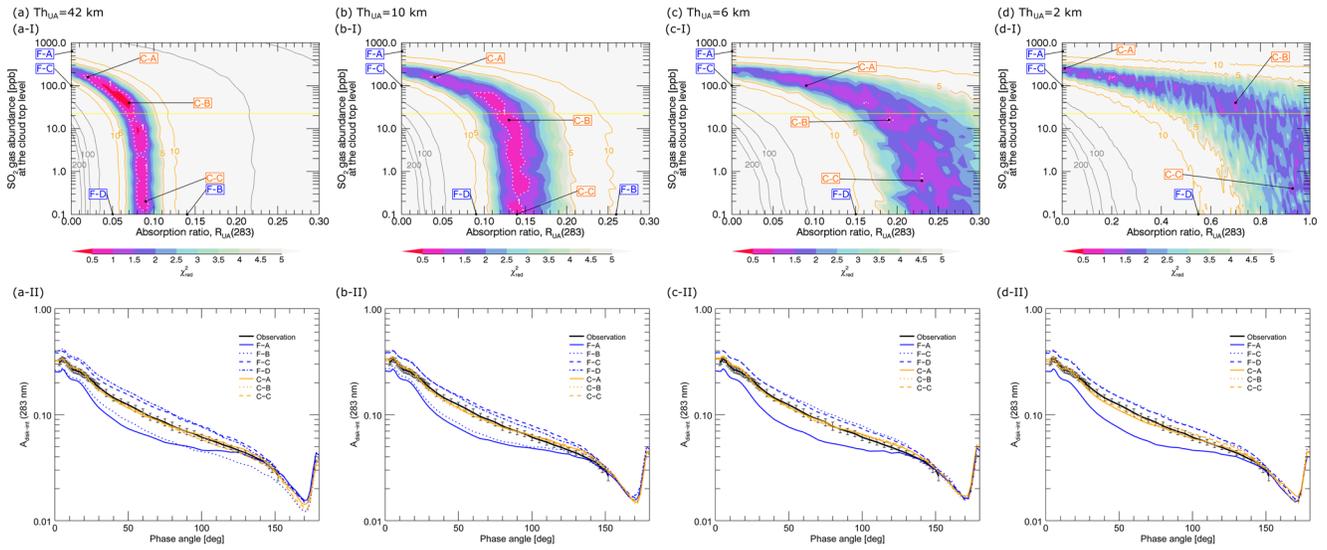

**Figure 3.** Comparison of the observed and calculated phase curves at 283 nm. $Th_{UA}$ varies from 42 (entire cloud) to 2 km, from left to right columns. $z_{UA-cld}$ adopts the best solutions at 365 nm (Panels [b–d], "×"). Other $z_{UA-cld}$ values are used in Figure S8. The upper panels show $\chi^2_{red}$ contours as a function of $R_{UA}(283)$ and $SO_2$ abundance at the cloud top. Note the range of $R_{UA}(283)$ in (d), which is upto 1.0. The yellow line refers to the reference $SO_2$ abundance, 22.4 ppbv (see the text for details). The white dotted lines indicate $\chi^2_{red}=1$. The area with the values beyond the range of the color bar ($\chi^2_{red} > 5$) is filled with light gray. Orange contours have a different interval than gray contours. The lower panels compare phase curves for selected cases as marked in the upper panels: false cases (from F-A to F-D), and comparably good solutions (from C-A to C-C). Their exact conditions and $\chi^2_{red}$ values are listed in Table S4.

For $Th_{UA} = 6$ km (Figure 2c) the best solution corresponds to $R_{UA}(365) = 0.09$ and $z_{UA-cld} = -3$ km ($\chi^2_{red} = 0.19$). The $\chi^2_{red} \leq 1$ area indicates that $z_{UA-cld}$ should be between $-1$ and $-8$ km. At $z_{UA-cld} = -8$ km (UA at 59–65 km), the sensitivity to $R_{UA}(365)$ becomes weak even for $R_{UA}(365) > 0.10$, as the UA layer reduces its impact on the planet brightness due to its deep location.

For $Th_{UA} = 2$ km (Figure 2d) the best solution corresponds to $R_{UA}(365) = 0.32$ and $z_{UA-cld} = -3$ km ($\chi^2_{red} = 0.12$). The area of $\chi^2_{red} \leq 1$ indicates that $z_{UA-cld}$ should be between $-1$ and $-6$ km. At $z_{UA-cld} = -6$ km (UA at 63–65 km), the sensitivity to $R_{UA}(365)$ becomes weak even for $R_{UA}(365) > 0.40$, for the reasons discussed above.

All our analyses at 365 nm lead to one conclusion: UA absorption within the 5 km below the cloud top is necessary to explain the observed phase curve.

### 4.2. 283 nm Results

Next, we investigate the 283 nm phase curves. As before, $Th_{UA}$ varies from 42 to 2 km. For $Th_{UA} = 2$–10 km, we vary $z_{UA-cld}$ from $-1$ to $-5$ that is within the range of good solutions at 365 nm. All results with varying $z_{UA-cld}$ and $Th_{UA}$ are shown in Figure S8, and Figure 3 shows the results with a fixed $z_{UA-cld} = -3$ km, which is the common best solution at 365 nm.

In Figure 3, for each $Th_{UA}$, the free parameters are $R_{UA}(283)$ and $SO_2$. The $\chi^2_{red}$ contours are shown in the upper panels (Figure 3I). Selected phase curves are compared in the lower panels (Figure 3II). A degeneracy is seen between the absorption introduced by the UA and $SO_2$ at this wavelength; the $SO_2$ abundance can be any value below 200–300 ppbv, strongly depending on $R_{UA}(283)$. Similar conclusions are found for the varying $z_{UA-cld}$ (Figure S8).

## 5. Discussion

The observed phase curves result from scattering, mainly by the cloud aerosols, and absorption by $SO_2$ and the UA. As the scattering by the cloud aerosols has a strong phase angle dependence, the effective absorption varies with phase angles especially when it occurs below the cloud top level. In Section 4.1, we showed that this is the case for the UA at 365 nm. We find that the UA could exist in the entire upper cloud or in a thin layer in the cloud, but the necessary condition for all UA scenarios is that sufficient absorption occurs within a few kilometers immediately below the cloud top. So, for any finite depth of UA layer ($Th_{UA}$ = 2–10 km), we find that it must be located near but below the cloud top level. In terms of our model parameters, this involves that the top of the UA layer ($z_{UA,top}$) should be, at the deepest, ∼5 km below the cloud top (Figure S6a and C-B in Figures 2c–2d). Otherwise, the scattering above $z_{UA,top}$ hides the UA absorption (e.g., F-C and F-D in Figure 2).

For all finite UA layers, all of C-B in Figures 2b–2d are possible solutions at 365 nm, and all of these have $z_{UA,top}$ = −5 km. However, such UA layer with $z_{UA,top}$ = −5 km produces negligible absorption at 283 nm that is muted in the disk-integrated signal by increased Rayleigh scattering at the shorter wavelengths (2.8 times stronger at 283 nm than at 365 nm) (Figure S7). Such a scenario cannot explain the observed $SO_2$ abundance that will be discussed later. A similar behavior is also shown in Figure S8cIII. These findings imply that to satisfy the brightness requirements at the two wavelengths, sufficient absorption of the UA should be within one cloud scale height immediately below the cloud top, and that a better fit at 283 nm is achieved when the UA layer is closer to the cloud top level (Figure S8I). We additionally find that $z_{UA,bot}$ < −10 km (for all other parameters remaining the same) does not affect the phase curves (Figure S6b), implying that our model is only sensitive to altitudes above 60 km (the upper cloud layer).

With the possible solutions of the UA layer at 365 nm, we attempted to constrain the $SO_2$ abundance and $R_{UA}$(283). However, we find that the phase angle dependence is insufficient for such a distinction (Section 4.2), and that both properties are degenerate in our model even though their prescribed vertical distributions are different. To gain insight into this degeneracy, we compare our results with the temporal mean of global $SO_2$ abundance in years 2016–2018, retrieved from midinfrared observations (Encrenaz et al., 2019); the averaged 450 ppbv at 61 km altitude in 2016–2018 translates into 22.4 ppbv at 70 km with the 3 km scale height of $SO_2$ abundance distribution. Such a $SO_2$ abundance at 70 km in years 2006–2014 showed a successful consistency between the existent midinfrared and UV observations (Figure 18 in Encrenaz et al. (2019) and Figure 7 in Marcq et al., 2020). So, we expect the same consistency in 2016–2018, and the 22.4 ppbv $SO_2$ is our reference. By doing so, we find the required $R_{UA}$(283) as 0.07–0.09 ($Th_{UA}$ = 42 km), 0.13–0.14 ($Th_{UA}$ = 10 km), 0.19–0.23 ($Th_{UA}$ = 6 km), and >0.8 ($Th_{UA}$ = 2 km), thereby corresponding to C-B in Figure 3. These $R_{UA}$(283) are more than twice $R_{UA}$(365) for the same $Th_{UA}$ and $z_{UA-cld}$ (Tables S3–S4), compensating the increased Rayleigh scattering at 283 nm. The required $R_{UA}$(283) is slightly lower with $z_{UA-cld}$ = −1 km (Figure S8I) than those with $z_{UA-cld}$ = −3 km (Figure 3), but still $R_{UA}$(283) > $R_{UA}$(365) for all cases. The possible impact of the high-latitudinal ozone would not be an important source of error; its global mean abundance is ∼5% of the $SO_2$ reference, while their absorption cross sections are comparable at 283 nm (Marcq et al., 2020).

If the increasing absorption of UA from 365 to 283 nm is real, then it would hint at possible UA candidates. We find a similar wavelength dependence of absorption from disulfur dioxide (OSSO) (Wu et al., 2018) and the normal $S_2O$ isomer (Frandsen et al., 2020). Wu et al. (2018) measured the absorbance of four types of OSSO, and showed that it peaks at 287 nm and is therefore stronger than at 365 nm. Frandsen et al. (2020) calculated the absorption cross section of the normal $S_2O$ isomer, and showed its peak is near 283 nm, overlapping with the $SO_2$ absorption peak. Interestingly, these are the two best solutions in Pérez-Hoyos et al. (2018), although the authors used different absorption spectra for OSSO and $S_2O$ from those in Wu et al. (2018) and Frandsen et al. (2020), respectively. However, our findings on the nature of the UA should be seen as tentative, since we have not explored other possibilities, such as embedding the UA only in mode 1 or mode 2 (but not in both) or different sizes of mode 1 particles. Future analyses with spectral data are necessary to constrain better the UA candidates. We are currently exploring the spectral disk-integrated albedo measurements obtained from the BepiColombo spacecraft during its cruise toward Mercury (Lee, Helbert, et al., 2019). Spectroscopy may allow us to investigate the degeneracy between $SO_2$ and the UA at nearby wavelengths in the vicinity of 283 nm that are sensitive to different extents to $SO_2$ absorption.

Our phase curve analysis assumes global mean conditions, and we do not investigate latitudinal or local time variations (e.g., the observed latitudinal variability of the 365 nm brightness is described in Lee et al., 2020). Averaging over 3 years of disk-integrated data should remove most of the variability across the planetary disk. We do not analyze detailed microphysical properties of cloud aerosols, as the current mode 2 size (Lee et al., 2017) explains the glory feature well. Our results provide plausible configurations for the vertical distribution of UA that can be further explored especially for the solar energy deposition calculation which has a considerable impact on the radiative energy balance and the global circulations in the Venusian atmosphere (Lee, Jessup, et al., 2019).

## 6. Conclusion

We analyzed the disk-integrated phase curves of Venus at 283 and 365 nm in the 3°–152° phase angle range. The observational data over 3 years were acquired by the UV camera on board Akatsuki. We calculated synthetic phase curves with a backward Monte Carlo algorithm to take into account multiple scattering over the full-disk. We explored the 365 nm phase curves with models that incorporate a number of free parameters on the unknown absorber, namely: vertical location and thickness of the unknown absorber layer, and absorption coefficient of the unknown absorber. Our results show that the unknown absorber can exist in the entire upper cloud, or as a thin layer, but it is required that sufficient absorption be present within one cloud scale height immediately below the cloud top. From our analysis at 283 nm, it turns out that we cannot separate absorption by $SO_2$ and the unknown absorber. Comparing our results with the $SO_2$ abundance retrieved from the midinfrared observations, we arrive at the tentative conclusion that the absorption coefficient of the unknown absorber at 283 nm is more than twice that at 365 nm. This increased absorption at 283 nm compensates the increased Rayleigh scattering at the same altitudes.

## Data Availability Statement

UVI data are publicly available at the JAXA archive website, DARTS (http://darts.isas.jaxa.jp/), and the NASA archive website, PDS (https://pds.nasa.gov/). UVI level 3x products are used in this study, L3bx (doi:10.17597/ISAS.DARTS/VCO-00016, Murakami et al., 2018), and the mean phase curve data are tabulated in supplementary Table S2. Processed individual disk-integrated albedo values at 283 and 365 nm are available as supplementary data of Lee et al. (2020).

**Acknowledgments**
The PBMC algorithm is available upon request to the code author, A. García Muñoz. Y. J. Lee has received funding from EU Horizon 2020 MSCA-IF No. 841432. Open access funding enabled and organized by Projekt DEAL.